\documentclass[12pt,a4paper]{iopart}

\usepackage{graphicx}
\usepackage{iopams}
\usepackage{xspace}

\newcommand{\sgn}{\ensuremath{\mathrm{sgn}}\xspace}

\newcommand{\dvec}[1]{\ensuremath{\boldsymbol{#1}}\xspace}
\newcommand{\vd}{\dvec{\mathrm{d}}}
\newcommand{\vk}{\dvec{\mathrm{k}}}
\newcommand{\vr}{\dvec{\mathrm{r}}}

\newcommand{\abs}[1]{\ensuremath{\left| #1 \right|}}

\begin{document}

\title[Interface states in Dirac materials]%
	{The role of spin-orbit coupling in topologically protected
	interface states in Dirac materials}
\author{D.\,S.\,L.\,Abergel$^1$}
\author{Jonathan M.\,Edge$^1$}
\author{Alexander V.\,Balatsky$^{1,2}$}
\address{$^1$ Nordita, KTH Royal Institute of Technology and Stockholm
University, Roslagstullsbacken 23, SE-10691 Stockholm, Sweden}
\address{$^2$ Los Alamos National Laboratory, Los Alamos, NM 87545, USA}

\pacs{73.20.-r, 73.40.-c}

\begin{abstract}
We highlight the fact that two-dimensional materials with Dirac-like
low energy band structures and spin-orbit coupling will produce 
linearly dispersing topologically protected Jackiw-Rebbi modes at
interfaces where the Dirac mass changes sign. These modes may support
persistent spin or valley currents parallel to the interface, and the
exact arrangement of such topologically protected currents depends
crucially on the details of the spin-orbit coupling in the material. As
examples, we discuss buckled two-dimensional hexagonal lattices such as
silicene or germanene, and transition metal dichalcogenides such as
MoS$_2$.

\noindent{\it Keywords\/}: Dirac materials, interface states, persistent
currents, topological protection.
\end{abstract}

\maketitle

\section{Introduction}

A major issue in contemporary condensed matter physics is the attempt to
produce and control currents which are polarized in one or more of
several spin-like degrees of freedom, and which can be used for
switching and possibly quantum computing applications. 
The real electron spin was initially proposed for this, leading to the
moniker ``spintronics''
\cite{Zutic-RMP76} and many advances in commercially relevant
technologies.
Recently, the isolation of two-dimensional (2D) hexagonal crystals with
their two inequivalent valleys and two inequivalent lattice sites have
lead to the suggestion of using these degrees of freedom in an analogous
way in so called ``valleytronics'' \cite{Rycerz-NatPhys3} and
``pseudospintronics'' \cite{Pesin-NatMat11} applications.
In general, one way to create polarization in one of these spin-like
degrees of freedom is to break a symmetry of a system such
that the degeneracy of the two spin-like flavors is lifted and
transport is only permitted for one flavor, for example using the giant
magneto-resistance effect \cite{Prinz-Science282}.
Another way is to utilize differences in how each flavor behaves
under an external perturbation \cite{Abergel-APL95}.
A third way is to employ the topological properties of materials in
order to engineer spatially localized states which can be manipulated
via the topology of the underlying material.
One such example of this is the 2D surface state which
forms at the interface between bulk Bi$_2$Se$_3$ crystals and the
vacuum \cite{Xia-NatPhys5, Zhang-NatPhys5-TI},
where the change in the $Z_2$ invariant from the crystal to the
vacuum implies that there must be a local closing of the band gap and
an associated interface mode. 
These interface modes are protected against disorder because the change
in the topological nature of the material on either
side of the interface requires the presence of such metallic surface
states.

In this article, we investigate the details of a localized and
controllable version of the phenomenon of current polarization enforced
by topological change in Dirac materials \cite{Wehling-AdvPhysSub}.
Localized interface modes protected by topological changes induced by a
change in sign of the band mass at the boundary of a system are known as
``Jackiw-Rebbi modes'' \cite{Jackiw-PRD13}. \footnote{Other authors have
labeled these modes as ``topologically confined modes'' and
``topological zero line mode''.}
These interface modes support persistent currents which can, in
principle, be utilized for electronic control and switching
applications, and may lead to significant efficiency improvements over
conventional charge switching technology \cite{Qiao-arXiv}. 
Bilayer graphene was considered in a similar context
\cite{Martin-PRL100, Li-PRB82}, and complementary results were shown,
including proposals to manipulate such states in ``electronic highways''
\cite{Qiao-NanoLett11}. Also, such modes can be hosted by the staggered
sublattice potential associated with the interface between graphene and
hexagonal boron nitride \cite{Jung-NanoLett12}, and in certain
perovskite materials \cite{Liang-NJP15}.

Our analysis corresponds to a different class of system where the
underlying band structure is governed by the linear Dirac Hamiltonian,
and we wish to highlight a fundamentally different feature, namely that
the details of the spin-orbit coupling (SOC) play a crucial role in
determining specific properties of the interface modes.
In particular, the variety of behavior seen in different materials 
caused by the interplay between the SOC and other effects related to the
lattice gives a handle for selecting the properties of interface modes
for specific applications.

For clarity, in the remainder of this introduction we demonstrate that
in the generic case where there is a band gap at the Fermi energy and a
finite quasiparticle mass, arranging a system such that there is an
interface where the Dirac mass is positive on one side and negative on
the other will ensure that topologically protected modes are present at
the interface.
Then, in Sec.~\ref{sec:realmaterials} we we shall show that the
persistent spin or valley currents carried by these interface modes
depend on the detailed nature of the SOC and the
interplay between the SOC and the sublattice asymmetry in each specific
material.  Section~\ref{sec:conclusion} contains our conclusions and
some discussion.

To illustrate the general point of topological modes confined by mass
inversion, the Hamiltonian for the bulk of a generic hexagonal lattice
can be written for a single valley and spin as
\begin{equation}
	H = \hbar v( \hat{k}_x \tau_x + \hat{k}_y \tau_y ) + \Delta \tau_z,
	\label{eq:genham}
\end{equation}
where $\Delta$ is the Dirac mass, $\hat{k}_{x} = -i\partial_x$ and
$\hat{k}_y = -i\partial_y$ are the wave vector operators, $v$ is the
Fermi velocity, and $\tau_{x,y,z}$ are Pauli matrices in the sublattice
space.
We begin by demonstrating that changing the sign of the band mass
$\Delta \to -\Delta$ in the bulk will alter the topological properties
of this toy system.
The topological quantum number associated with a Hamiltonian such as
Eq.~\eref{eq:genham} is the Chern number $C(\Delta)$ \cite{Ryu-NJP12}.
It may be expressed as an integral of the Berry curvature over the
Brillouin zone and for a system described by $H = \sum_{i} d_i(\dvec{k})
\tau_i$ [the Hamiltonian in Eq.~\eref{eq:genham} has $\vd = (\hbar
v\hat{k}_x, \hbar v \hat{k}_y, \Delta)$], the Chern number can be
calculated from \cite{Thouless-PRL49, Qi-PRB78}
\begin{eqnarray}
   C(\Delta)&=\frac1{4\pi} \left(
		\int_0^\Lambda d^2 \vk + \int_{\abs{k}>\Lambda} d^2 \vk \right)
		\epsilon^{\mu\nu\lambda} \bar{d}_{\mu}\partial_{k_x}
		\bar{d}_{\nu}\partial_{k_y} \bar{d}_{\lambda} \nonumber\\
	&=C^{(1)}(\Delta,\Lambda) + C^{(2)}(\Delta,\Lambda).
	\label{eq:def_chern_no}
\end{eqnarray}
In this equation, we have split the integration into two contributions.
The first, $C^{(1)}$, comes from wave vectors near to the gapless point,
and the second, $C^{(2)}$, from the rest of the Brillouin zone where we
assume there are no further gap closings.
A cut-off $\Lambda$ distinguishes these two regions,
$\bar{\vd}= \vd/\abs{\vd}$, and $\epsilon$ is the three-dimensional
Levi-Civita symbol.
We examine the change in the Chern number $\delta C$ when the Dirac mass
is inverted by calculating $\delta C = C(\Delta) - C(-\Delta)$ so that a
finite value of $\delta C$ indicates a change in the system's
topological properties.
We know that the region of $k$-space near the gapless point is the only
place where the conduction and valence bands come near to each other,
and that the gapped part of a Hamiltonian cannot introduce a change in
the Chern number. Hence, $\delta C$ may be computed solely from the
low-momentum part of Eq.~\eref{eq:def_chern_no}.
Hence, $\delta C = C^{(1)}(\Delta) - C^{(1)} (-\Delta)$ and, in the limit
$\Delta \to 0$ and $\Lambda \gg \Delta/(\hbar v)$, we obtain
\begin{equation}
	\delta C = \frac{1}{4\pi} \int_0^{\Lambda}
	\frac{2\hbar^2 v^2\Delta }{(\hbar^2 v^2\abs{\vk}^2 + \Delta^2)^{3/2}}
	d^2k = 1.
	\label{eq:deltaC}
\end{equation}
Therefore, since there is a change in the Chern number when the sign of
$\Delta$ is reversed, the boundary between regions of a system with
$\Delta>0 $ and $\Delta<0 $ hosts a topologically protected interface
state.
\footnote{Note that there is some subtlety associated with this
analysis, as described in Ref.~\cite{Li-PRB82}. But the change in
Chern number across a domain wall is a well-defined topological
invariant in the cases we consider.}

We now demonstrate the localization of a linear ``interface mode'' in the
region where the band mass changes sign. This corresponds exactly to the
Jackiw-Rebbi modes introduced earlier \cite{Jackiw-PRD13}.
We allow the mass term to become inhomogenous, such that it is constant
in the $y$ direction, but has linear slope in the $x$ direction with
$\Delta(0)=0$ then $\Delta(\vr) = \Delta_0 x/R$ so that $1/R$
characterises the gradient of the change in the mass and a
one-dimensional domain wall is defined.
The spectrum of this system is found by rotating the Hamiltonian
with the unitary operator $U=e^{i\pi \tau_x/4}$.
Since $\hat{k}_y$ commutes with the Hamiltonian, we replace it with the
eigenvalue $k_y$ and hence $H' = -i \hbar v \partial_x \tau_x + \hbar v
k_y \tau_z - \Delta_0 x \tau_y / R$.
The off-diagonal elements can now be written in the form of ladder
operators $a = \sqrt{\alpha/2}( x + \partial_x/\alpha)$ and
$a^\dagger = \sqrt{\alpha/2}(x - \partial_x/\alpha)$ with
$\alpha = \Delta_0 / (\hbar vR)$ which act on the harmonic oscillator
functions $\Phi_n$ as $a^\dagger \Phi_n = \sqrt{n+1}\Phi_{n+1}$, $a
\Phi_n = \sqrt{n} \Phi_{n-1}$, and $a \Phi_0 = 0$.
We can construct by inspection an eigenvector of $H'$ which has the form
\begin{equation*}
	\Psi = e^{ik_y y} \left( 
		\begin{array}{c}\Phi_0 \\ 0 \end{array} \right)
\end{equation*}
and dispersion $\varepsilon = \hbar v k_y$.
This mode behaves as zero gap semiconductor despite the mass gap in the
bulk system, and is topologically protected by the change in the Chern
number across the domain wall.

\begin{figure}[tb]
	\centering
	\includegraphics[]{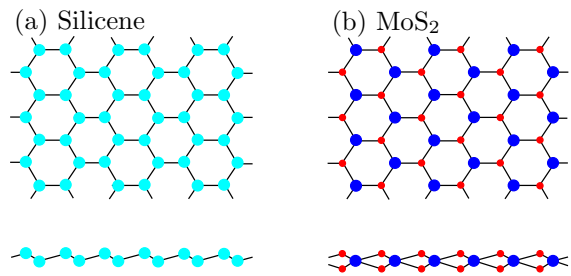}
	\caption{Lattice structures of (a) silicene, and (b) molybdenum
	disulphide, showing both a top-down view (upper sketch) and side-on
	view (lower sketch).
	\label{fig:lattices}}
\end{figure}

\section{Real systems \label{sec:realmaterials}}
The precise details of the manifestation of the topologically protected
interface states will depend on the microscopic characteristics of the
material in which they are realized, so we give two examples of physical
systems which fulfil the requirements outlined above.

\subsection{Silicene}
The first, silicene, has already been discussed in the context of
interface modes \cite{Ezawa-NJP14, Ezawa-PRB87}, although valley
currents were not mentioned.
Theoretical proposals for transport through normal-ferromagnetic-normal
junctions \cite{Yokoyama-PRB87} and gated regions \cite{Tsai-NatCommun4}
which produce spin or valley currents have been reported, but this is a
fundamentally different mechanism from the interface state transport we
discuss.
For silicene, the $\tau_z$ term in the Hamiltonian represents the
combination of the SOC and the asymmetry between the on-site potential
of the two sublattices in the hexagonal crystal.
The SOC term enters with the $\tau_z$ Pauli matrix
because the warping of the lattice which generates the coupling is
opposite on the two sublattices \cite{Liu-PRB84}.
Also, because silicene exhibits a buckled structure [as illustrated in
Fig.~\ref{fig:lattices}(a)], applying a transverse electric field
adjusts the static potential on the two sublattices in different
magnitude, hence modifying the mass.
Silicene is known to have a rich phase diagram in the non-interacting
regime \cite{Ezawa-PRB87} in
which varying the transverse electric field causes the 2D bulk to
undergo a phase transition from a quantum anomalous Hall phase
to a band insulator phase due to the change in the Dirac mass.
In common with all materials that have a 2D hexagonal lattice, silicene
exhibits the valley pseudospin in addition to the real
electron spin and lattice pseudospin.
The low-energy effective Hamiltonian for electrons of spin $s=\pm 1$ in
valley $\xi=\pm 1$ of silicene is \cite{Liu-PRB84}
\begin{equation}
	H_{\xi s}^{\mathrm{Sil}} =
	\hbar v( \hat{k}_x \tau_x - \xi \hat{k}_y \tau_y )
	+ \xi s \lambda \tau_z + \frac{l E_z}{2} \tau_z
	\label{eq:silham}
\end{equation}
where $v$ is the Fermi velocity associated with the Dirac spectrum, the
Pauli matrices $\tau_{x,y,z}$ are in the sublattice space, $\lambda$
parameterizes the strength of the SOC, $l E_z$ is the
contribution to the band gap induced by external gating, and $\hat{k}_i
= -i\hbar \partial_{x_i}$ is the operator for the electron wave vector.
The inhomogeneous electric field is assumed to define an interface
oriented along the $y$ axis at $x=0$, and is represented as $E_z =
x\mathcal{E}/R$.
The effective mass of the Dirac fermions is negative when
$lE_z(x) < - \xi s \lambda$, and positive otherwise.
Therefore, for bands with $\xi s = 1$, there are
interface modes confined near $x=-\lambda R/(\mathcal{E}l)$ and a
further two modes with $\xi s = -1$ near $x=\lambda R/(\mathcal{E}l)$.
This is a difference from our toy example discussed above, where the
band was localized near $x=0$, and is due to the combination of the
intrinsic SOC in silicene and the electric field in
creating the total Dirac mass.
To find the dispersion of these interface modes, we apply the same
series of manipulations as in our toy system which yields
$\varepsilon^{\mathrm{Sil}}_{\xi s} = -\xi \hbar v k_y$.

\begin{figure}[bt]
	\centering
	\includegraphics[]{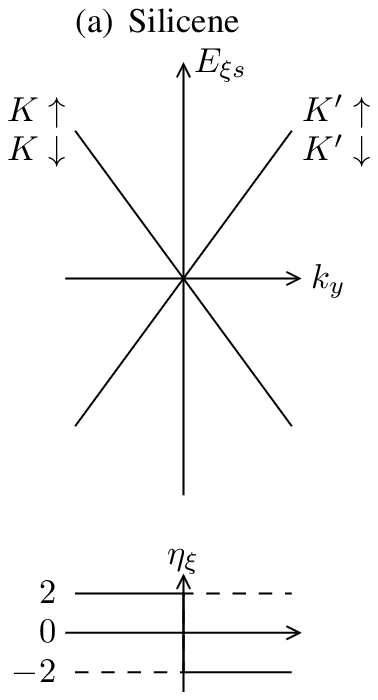}%
	\hspace{2cm}%
	\includegraphics[]{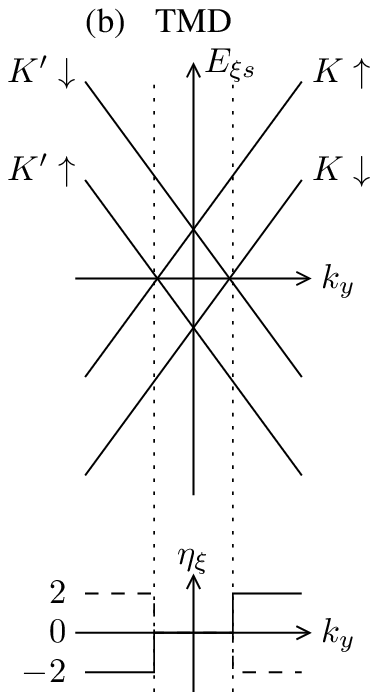}
	\caption{(a) The topologically protected interface modes in
	silicene. The two spin states in each valley are degenerate.
	(b) The topologically protected bands in TMDs.
	The lower plots in (a) and (b) show the valley band asymmetry
	parameter $\eta_\xi(k_y)$ for $\xi=K$ (solid lines) and $\xi=K'$
	(dashed lines).
	\label{fig:interfacemodes}}
\end{figure}

We can use a topological characteristic to describe the spectral
asymmetry of the Hamiltonian which is enforced by the topological modes.
We define the indices \cite{Balatsky-JETPLett47}
\begin{eqnarray}
	\eta_s = \frac{2}{\sqrt{\pi}} \Tr \int_0^\infty H e^{-y^2 H^2} P_s
	dy \\
	\eta_\xi = \frac{2}{\sqrt{\pi}} \Tr \int_0^\infty
		H e^{-y^2 H^2} P_\xi dy
\end{eqnarray}
where $P_s$ and $P_\xi$ are, respectively, the projection operators onto
spin $s$ and valley $\xi$. The notation $\Tr$ denotes a sum over all spins
and valleys.
This index counts the excess number of bands with positive energy.
As shown in Fig.~\ref{fig:interfacemodes}(a), applying this definition
to the topological bands of the Hamiltonian in
Eq.~\eref{eq:silham} gives $\eta_{K} = 2\mathrm{sgn}(k_y)$ and
$\eta_{K'} = -2\sgn(k_y)$ because the electrons in valley $K$ and in
valley $K'$ have opposite sign of their energy for a given $k_y$. This
index indicates that there is a fundamental asymmetry in the valley
distribution of the interface modes in silicene.
Conversely, the spin characteristic is zero because of the degeneracy
of the bands.

We can also describe one-dimensional spin or valley currents along the
interface in terms of the group velocity of electrons in each band. For a
single band,
\begin{equation*}
	j_{\xi s} = \frac{1}{2\hbar k_c} \int_{-k_c}^{k_c}
	dk_y v_{\xi s} n_{\xi s}
\end{equation*}
where $v_{\xi s} = d\varepsilon^{\mathrm{Sil}}_{\xi s}/dk_y$,
$n_{\xi s}(k_y)$ is the occupation of the state with wave vector $k_y$
in the band with indices
$\xi$ and $s$, and $k_c$ is a cutoff wave vector of the order of the
Brillouin zone size.
Then, the total spin and total valley currents including the
contributions from all topological bands are
\begin{eqnarray*}
	j_s = j_{K\uparrow} + j_{K'\uparrow}
		- j_{K\downarrow} - j_{K'\downarrow}, \\
	j_\xi = j_{K\uparrow} + j_{K\downarrow}
		- j_{K'\uparrow} - j_{K'\downarrow}.
\end{eqnarray*}
where we have adopted the convention that $\uparrow$-spin ($K$ valley)
electrons moving in the positive $y$ direction give a positive
contribution to the total spin (valley) current.
In the case of silicene, the $K$ valley electrons are all left-moving
(with $v_{Ks} = -v$),
while the $K'$ electrons are all right moving (with $v_{K's} = v$),
clearly indicating that there is a valley current $j_\xi = -2v$ at
half filling.
The sign of this current can be reversed by inverting the gradient of
the electric field.
The higher energy bands are quadratic and degenerate in spin and valley
so do not contribute to the current.
If a finite chemical potential $\mu$ is introduced so that the system is
shifted away from half filling, a correction of $-2\mu/(\hbar k_c)$ is
added to the current. Since $\hbar v k_c \gg \mu$ by definition, this
effect is small indicating that the valley current is robust against
realistic changes in the density.

Silicene has been experimentally isolated \cite{Vogt-PRL108}, but the
disadvantage of this material is that the SOC is
relatively small ($\lambda \approx 4\mathrm{meV}$).
The SOC in a buckled lattice made from germanium is an order of
magnitude larger \cite{Liu-PRB84}, and for the interface modes we are
discussing, stronger SOC has the advantage of allowing a wider intrinsic
gap between the bulk bands. Therefore, the interface modes we discuss
would be easier to detect in that material.

\subsection{Transition metal dichalcogenides}
The second system we present is transition metal dichalcogenides (TMDs)
such as MoS$_2$. These materials are constructed of a hexagonal
bipartite lattice where a sheet of transition metal atoms on the A
sublattice are surrounded by two sheets of chalcogen atoms on the B
sublattice, as shown in Fig.~\ref{fig:lattices}(b).
The SOC comes primarily through the interaction with the
heavier transition metal atom \cite{Kormanyos-PRB88, Rostami-PRB88}, and
so the coupling is asymmetric in the lattice sites, in contrast to the
case of silicene. Hence we expect that there may be differences in the
manifestation of the topological modes.
The low-energy effective Hamiltonian in the bulk is \cite{Xiao-PRL108}
\begin{equation}
	H_{\xi s}^{\mathrm{TMD}} = \hbar v( \xi k_x \tau_x + k_y \tau_y ) +
	\frac{\Delta}{2} \tau_z - \frac{\xi s \lambda}{2} ( \tau_z - \tau_0 ).
\end{equation}
where the band mass $\Delta \approx 1.6\mathrm{eV}$ is a parameter of
the lattice coming from the intrinsic chemical asymmetry between the two
sublattices.
The band mass is assumed to vary in space, with the same
one-dimensional form $\Delta(x)=\Delta_0 x/R$ as was taken in the silicene
case.\footnote{However, we cannot suggest a physical mechanism for
achieving this spatial variation.}
The rotation of the Hamiltonian distributes the SOC term
over $\tau_y$ and $\tau_0$ so that the parameter $\lambda$ enters the
dispersion of the interface modes as
$\varepsilon^{\mathrm{TMD}}_{\xi s} = \xi \hbar v k_y + \xi s \lambda/2$.
This indicates that the band spin asymmetry index is $\eta_s = 0$
for all $k_y$, but that the valley index $\eta_\xi$ is finite for $|k_y| >
\lambda/(2\hbar v)$ [see Fig.~\ref{fig:interfacemodes}(b)].
This illustrates the fact that the arrangement of the interface modes is
different from silicene because of the different nature of the
SOC. A finite valley current exists with,
$j_\xi = 2 v + 2\mu/(\hbar k_c)$ which can also be reversed by switching
the sign of the gradient of the inhomogenous band mass.
In contrast with silicene, we also have an overall spin current
with $j_s = - \lambda/(\hbar k_c)$ which is set by the magnitude of the
SOC. This current does not depend on the location of the
chemical potential.

\section{Conclusions \label{sec:conclusion}}
In conclusion, we have demonstrated that the design and control of
mass inversion via careful selection of the host material for its band
gap and SOC is a theoretically viable technique for the
creation and manipulation of topologically protected modes that may
carry spin or valley polarized currents.
In silicene, the interface modes carry valley current which can be
controlled by local gating, and (if a mechanism for band inversion can
be found) TMDs such as MoS$_2$ will exhibit both spin and valley
currents at interfaces.  The direction of these currents can be
manipulated by reversing the sign of the gradient of the mass
inhomogeneity.
The topological protection we discuss is robust so long as there is no
inter-valley scattering. This is because the derivation of
Eq.~\eref{eq:deltaC} assumes that there is only one gapless
point. However, for realistic systems with hexagonal lattice structure,
there is a gapless point at each of the six Brillouin zone corners
(\textit{i.e.} one gapless point in each valley). We note that if the
valleys are connected, the assumptions inherent in
Eq.~\eref{eq:deltaC} are not satisfied. Thus, short-range
scatterers such as lattice defects will be severely detrimental to the
existence of interface modes.

\ack
This work was supported by ERC DM-321031 and by Nordita.

\end{document}